\documentclass[11pt]{article}

\usepackage{amsmath,amssymb,graphicx}
\usepackage{hyperref}

\usepackage{multicol,color,longtable}

\definecolor{darkred}{rgb}{0.65,0.15,0}
\hypersetup{pdfborder={0 0 0},colorlinks=true,urlcolor=darkred,citecolor=blue,linkcolor=darkred,linktocpage=true}

\newcommand{\nn}{\nonumber}

\newcommand{\reals}{\mathbb{R}}

\newcommand{\lb}{\left[}
\newcommand{\rb}{\right]}

\makeatletter

\@addtoreset{equation}{section}
\makeatother

\begin{document}

\thispagestyle{empty}
\mbox{ }
\vspace{20mm}

\begin{center}
{\LARGE \bf Enhancement of hidden symmetries\\[2mm] and Chern--Simons couplings}\footnote{To appear in the Proceedings  of the 9th Workshop and School on ``Quantum Field Theory and Hamiltonian Systems", 24-28 September 2014,  Sinaia, Romania.}\\[10mm]

\vspace{8mm}
\normalsize
{\large  Marc Henneaux${}^{1,2,3}$, Axel Kleinschmidt${}^{3,4}$ and Victor Lekeu${}^{2,3}$}

\vspace{10mm}
${}^1$ {\it Centro de Estudios Cient\'ificos (CECS), Casilla 1469, Valdivia, Chile}
\vskip1em
${}^2${\it Universit\'e Libre de Bruxelles, ULB-Campus Plaine CP231, B-1050 Brussels, Belgium}
\vskip1em
${}^3${\it International Solvay Institutes\\
ULB-Campus Plaine CP231, BE-1050 Brussels, Belgium}
\vskip 1 em
${}^4${\it Max-Planck-Institut f\"{u}r Gravitationsphysik (Albert-Einstein-Institut)\\
Am M\"{u}hlenberg 1, DE-14476 Potsdam, Germany}
\vskip 1 em
\vspace{5mm}

%\hrule

\vspace{10mm}

\begin{tabular}{p{12cm}}
{\small
We study the role of Chern--Simons couplings for the appearance of enhanced symmetries of Cremmer--Julia type in various theories. It is shown explicitly that for generic values of the Chern--Simons coupling there is only a parabolic Lie subgroup of symmetries after reduction to three space-time dimensions but that this parabolic Lie group gets enhanced to the full and larger Cremmer--Julia Lie group of hidden symmetries if the coupling takes a specific value. This is heralded by an enhanced isotropy group of the metric on the scalar manifold.  Examples of this phenomenon are discussed as well as the relation to supersymmetry. Our results are also connected with rigidity theorems of Borel-like algebras.
}
\end{tabular}
\vspace{10mm}
%\hrule
\end{center}

\newpage
\setcounter{page}{1}

%\tableofcontents

\section{Introduction}

Gravitational theories coupled to $p$-form fields can exhibit remarkable ``hidden" symmetries of Cremmer--Julia type \cite{Cremmer:1979up} when the Chern--Simons couplings among the $p$-forms take specific values.  The purpose of this note is to investigate in detail the dependence of these symmetries on the Chern-Simons couplings\footnote{We only consider here symmetries of Cremmer--Julia type.  There might be other symmetries but these will not concern us.}.  It has been observed previously that for generic values of the couplings, the theory is invariant only under a smaller algebra.   Even though the structure constants of this smaller symmetry algebra $A$ depend explicitly on the Chern--Simons coefficients, we show that these coefficients can be absorbed in the structure constants of $A$ through redefinitions of the basis of $A$ (except for a subset of  isolated values corresponding to contractions of the Lie algebra). This enables one to identify the smaller symmetry algebra with the parabolic subalgebra of the full hidden symmetry algebra that appears for the critical values of the couplings. This property is related to rigidity theorems preventing non-trivial deformations of Borel-like Lie algebras \cite{LegerLuks}.

One motivation for undertaking this study is supersymmetry. It is common lore that the appearance of large hidden symmetries in supergravity is tantamount to the presence of supersymmetry.  To the best of our knowledge, however, an explicit demonstration of this fact has not appeared in the literature.  Our note  fills this gap.

The connection between hidden symmetries and supersymmetry of supergravity theories comes as follows. Supersymmetry fixes the value of the couplings between all the fields in a multiplet, including the self-couplings. The prime instance of this phenomenon can be seen in maximal supergravity in $D=11$ space-time dimensions where the gravity multiplet contains (on-shell) as bosonic fields the metric $g_{MN}$ and a three-form $A_{MNP}=A_{[MNP]}$ that are governed by the Lagrangian density~\cite{Cremmer:1978km}
\begin{align}
\label{SUGRA11}
\mathcal{L}_{(11)}  =  R \star 1\!\!1 - \frac12 \star F \wedge F + \frac16 F\wedge F\wedge A.
\end{align}
Here, we have employed form notation for the fields and $F=dA$ is the field strength four-form of $A$ that is invariant under gauge transformations $A\to A+ d\Lambda$ for any two-form $\Lambda$. The last self-interaction term is a Chern--Simons term and it varies into a total derivative under the gauge transformation. Its coupling coefficient $\frac16$ is fixed by supersymmetry if one constructs the full supergravity theory~\cite{Cremmer:1978km}. The value of the  Chern--Simons coupling is {\em not} fixed by gauge symmetry or diffeomorphism symmetry.

It is a celebrated feature of $D=11$ supergravity that it exhibits a chain of so-called hidden symmetries when it is dimensionally reduced. Reduction to $D=4$ results in the ungauged $\mathcal{N}=8$ supergravity theory with global $E_{7(7)}$ symmetry~\cite{Cremmer:1979up}.\footnote{This $E_{7(7)}$ symmetry is related to recently discovered improved UV finiteness properties of the theory~\cite{Bern:2011qn}, see~\cite{Beisert:2010jx,Kallosh:2011qt,Bossard:2011ij,Gunaydin:2013pma} for some discussions of full perturbative finiteness of the theory.}
Further reduction to $D=3$ results in a theory with only scalar propagating degrees of freedom encoded in the exceptional global symmetry $E_{8(8)}$, see~\cite{Julia:1980gr} where many similar cases are also discussed.

The relevance of $E_{8(8)}$ in the scalar sector can be seen from the kinetic terms of the theory using the formalism of~\cite{Lu:1995yn,Cremmer:1997ct,Lambert:2001he}. In this formalism one computes the so-called dilaton vectors of a dimensionally reduced theory where the kinetic terms of the axion fields $\chi$ are of the form $e^{\vec{\alpha}\cdot \vec{\phi}} (\partial \chi)^2$. The fields $\vec{\phi}$ are the dilatons that arise in the dimensionally reduced theory and $\vec{\alpha}$ are the dilaton vectors. If these can be identified with the {\em positive} roots of a Lie group one has thus identified a candidate hidden symmetry. We emphasise that this candidate hidden symmetry depends only on the kinetic terms of the theory and not on the Chern--Simons term. If one performs the analysis of the dilaton vectors for maximal supergravity one is directly led to $E_{8(8)}$~\cite{Cremmer:1997ct,Lambert:2001he}. 

Demonstrating the presence of the full non-linear $E_{8(8)}$, however, depends on the interaction terms of the theory and is therefore sensitive to the value of the Chern--Simons coupling in~\eqref{SUGRA11}. The full $E_{8(8)}$ also requires a role for all the {\em negative} roots of the symmetry and these are not automatically guaranteed by the symmetries of the kinetic terms. We will show that there is only an action of all negative step operators of the symmetry group if the Chern--Simons coupling has the right value and that this value is identical to the one required by supersymmetry.

\section{Global symmetries from dimensional reduction of gauge symmetries}

We begin by reviewing how some global symmetries arise from gauge symmetries in the process of dimensional reduction. This is lucidly explained in~\cite{Pope}.

\subsection{$GL(d,\reals)$ symmetry from diffeomorphisms}

Consider a theory in $D$ dimensions that is invariant under $D$-dimensional local diffeomorphisms. Dimensional reduction on a $d$-torus $T^d$ to $D-d$ dimensions restricts these diffeomorphisms. Throughout this note we take $d\leq D-3$. More precisely, writing the original $D$-dimensional coordinates as $x^M$ they are split into $x^M=(x^\mu, x^m)$ in the reduction procedure. Infinitesimal diffeomorphisms in $D$ dimensions are given by vector fields $\xi^M(x^N)$. For these to respect the reduction ansatz the `internal' components $\xi^m$ have to be of the special form
\begin{align}
\label{diffsplit}
\xi^m( x^\nu, x^n) = k^m{}_n x^n + \lambda^m(x^\nu)
\end{align}
with a {\em constant} $(d\times d)$-matrix $k^m{}_n$. Invertibility of diffeomorphisms restricts this matrix to lie in $GL(d,\reals)$.\footnote{The presence of the abelian diagonal $GL(1,\reals)$ part in $GL(d,\reals)\cong GL(1,\reals)\times SL(d,\reals)$ depends on the possibility of assigning appropriate scaling symmetries to the matter fields~\cite{Pope}; we will assume that this is possible for the theory under consideration.} This shows that the dimensionally reduced theory inherits a {\em global} $GL(d,\reals)$ from the originally local $D$-dimensional diffeomorphisms. (The parameter $\lambda^m(x^\mu)$ yields the gauge transformations of the Kaluza--Klein vectors arising in the reduction but is not of immediate importance to us since we will dualise all matter into scalars in three dimensions. It will, however, resurface later.) 

The global $GL(d,\reals)$ symmetries acts on the scalars in the lower $(d\times d)$ block $g_{mn}$ in the metric in the standard fashion. It also acts on all other fields in the theory as induced from the higher-dimensional diffeomorphism invariance.

\subsection{Shift symmetries from gauge fields}

Global symmetries in the reduced theory can similarly arise from other gauge symmetries present in the $D$-dimensional theory. Let us assume that the $D$-dimensional theory has a $p$-form field $A_{(p)}$ with gauge transformation $A_{(p)}\to A_{(p)} + d \Lambda_{(p-1)}$ under a $(p-1)$-form gauge parameter $\Lambda_{(p-1)}$. This if for example the case in $D=11$ supergravity displayed in~\eqref{SUGRA11} with $p=3$. In dimensional reduction one can obtain $(D-d)$-dimensional scalars from the $p$-form $A_{M_1\ldots M_p}$ whenever its indices are all internal. This only happens when $d\geq p$ and one obtains the scalars $A_{m_1\ldots m_p}$. The gauge parameters consistent with dimensional reduction are then of the form
\begin{align}
\Lambda_{m_1\ldots m_{p-1}} (x^\nu, x^n)= k_{m_1\ldots m_{p-1} n} x^n + \ldots
\end{align}
with a constant fully antisymmetric $k_{m_1\ldots m_p}= k_{[m_1\ldots m_p]}$. The dots denote additional terms that are functions of $x^\nu$ and that generate gauge transformations of non-scalar fields obtained in the dimensional reduction process. On the scalar components $A_{m_1\ldots m_p}$ the induced transformation is by constant shifts
\begin{align}
A_{m_1\ldots m_p} \to A_{m_1\ldots m_p} + k_{m_1\ldots m_p},
\end{align}
compatible with the fact that the scalars arising from the reduction of form fields are always of axionic type, i.e., they possess Peccei--Quinn shift symmetries. The transformation parameter $k_{m_1\ldots m_p}$ transforms under the $GL(d,\reals)$ that arose from the $D$-dimensional diffeomorphism symmetry since the original gauge parameter was a tensor. The relevant transformation is just the antisymmetric $p$-form representation of $GL(d,\reals)$. For this reason we obtain at this point a global symmetry of the type
\begin{align}
GL(d,\reals) \ltimes \reals^{N}\quad\textrm{with $N=\begin{pmatrix}d\\p\end{pmatrix}$.}
\end{align}

An important question now is what the structure of the shift symmetries is. If there are no other scalar fields in the theory, then the shifts in $\reals^N$ are just abelian, i.e. they commute. If there are other scalar fields one can obtain a more complicated structure of global shift symmetries. 

As a first example, we use $D=11$ supergravity. If one reduces this theory on a six-torus $T^6$ one generates $N=\frac{6\cdot5\cdot4}{3!}=20$ scalar fields $A_{m_1m_2m_3}$ from the completely internal components of the three-form gauge potential. However, there is also a completely `external' component $A_{\mu_1\mu_2\mu_3}$ that is still a three-form from the reduced five-dimensional perspective. However, one can perform a Hodge dualisation of the three-form to a scalar and there are therefore in total $21$ scalar fields arising from the eleven-dimensional gauge potential. A better way of saying this is that one can dualise the $D=11$ three-form to a six-form in $D=11$ and the extra scalar arises when all its six indices are internal. The new scalar field also comes with a shift symmetry so that there is in total a symmetry
\begin{align}
GL(d,\reals) \ltimes \left( \reals^{20} + \reals \right)
\end{align}
in five-dimensional supergravity. In this particular case, the structure of the shift symmetries is
\begin{align}
\label{336}
\lb k_{m_1 m_2 m_3} , k_{n_1n_2n_3} \rb = k_{m_1m_2m_3n_1n_2n_3},
\end{align}
where the fully antisymmetric $k_{m_1m_2m_3n_1n_2n_3}$ is the single shift generator associated with the single scalar that arose from the dualisation. This commutator is zero if there is no Chern-Simons interaction in the Lagangian. Additionally, the generator $k_{m_1m_2m_3n_1n_2n_3}$ commutes with $k_{m_1m_2m_3}$ and of course with itself.  We see that here we obtain a non-abelian group of shift symmetries from the original matter gauge symmetries but this depended on the original Lagrangian~\eqref{SUGRA11}. This example will be discussed in more detail in section~\ref{E8sec} below.

\subsection{General structure of global symmetries from gauge symmetries}

In general, the generators of all shift symmetries form a nilpotent global algebra. This means that forming commutators of a sufficiently high number of shift generators always yields zero. In the example~\eqref{336} above this is clear because any further commutator either with $k_{\ell_1\ell_2\ell_3}$ or with $k_{\ell_1\ldots \ell_6}$ gives zero. 

The general reason for this statement is that the shift symmetries are associated with matter fields of increasing form rank and the form rank is additive in the commutator algebra of shift transformations as exemplified in~\eqref{336}. Since the maximum rank is bounded by the number $d$ of internal directions one knows that taking multiple commutators will eventually lead to zero when one exceeds $d$. 

As a consequence we have that the general structure of the global symmetry obtained from gauge symmetries after dimensional reduction on a torus $T^d$ is
\begin{align}
\label{genG}
GL(d,\reals) \ltimes U
\end{align}
where $U$ is a unipotent group (the exponential of a nilpotent algebra) of the form
\begin{align}
\label{genU}
U=\sum_k U_k = U_1+U_2 +\ldots = r_1 \reals^{N_1} + r_2 \reals^{N_2} + \ldots
\end{align}
where 
\begin{align}
N_k = \begin{pmatrix} d\\k\end{pmatrix}
\end{align}
denotes the shift symmetries arising from $k$-form gauge symmetries and $r_k$ are potential degeneracies when there are multiple $k$-form gauge fields in the $D$-dimensional theory.  The commutators are such that they respect the additive grading by $k$ in~\eqref{genU} and there are only finitely many terms in the sum.

For generic values of the couplings, the global symmetries in~\eqref{genG} are the only Cremmer--Julia symmetries of the theory. They get enhanced, however,  when the couplings take specific values.

Before discussing the symmetry enhancement we have to make one more important comment. When one reduces a gravitational theory to three space-time dimensions (on a torus $T^{D-3}$) one obtains more scalar degrees of freedom from the dualisation of the Kaluza--Klein vectors in the metric. There are $D-3$ of these {\em dual graviton} scalars and their vector gauge symmetry (see~\eqref{diffsplit}) is also turned into a shift symmetry. This symmetry comes with a shift symmetry generator of the form
\begin{align}
k_{m_1\ldots m_{D-3}, n} \quad \textrm{such that}\quad k_{[m_1\ldots m_{D-3}],n} = k_{m_1\ldots m_{D-3},n}  \quad\textrm{and}\quad k_{[m_1\ldots m_{D-3},n]} =0. 
\end{align}
For $D=4$ Einstein gravity this was first observed by Ehlers~\cite{Ehlers} and the dual graviton has played a prominent role in recent discussions of conjectural infinite-dimensional symmetries~\cite{West:2001as,Damour:2002cu}. The presence of this additional shift symmetry does not invalidate the argument of $U$ being unipotent and the generator appears in the graded expansion~\eqref{genU} at degree $k={D-2}$, corresponding to its scaling weight.

The global symmetries in~\eqref{genG} (with the complete $U$ including all shift symmetries) form a parabolic subgroup of the enhanced symmetry group of Cremmer--Julia type.\footnote{By parabolic group we here mean a group of the form $L\ltimes U$ with $L$ reductive and $U$ unipotent. Normally, parabolic is defined for subgroups of some larger groups in which parabolic subgroups contain a Borel subgroup. In the case of symmetry enhancement this will be exactly the situation we have here.} This will be discussed below. For the case of $D=11$ supergravity this group is a maximal parabolic subgroup of $E_{8(8)}$ so we are still some way from having the presence of all of $E_{8(8)}$.

\subsection{Scalar manifold and its geometry}

The scalar fields of the dimensionally reduced theory parametrise a scalar manifold on which we can give an explicit choice of coordinates by
\begin{align}
V = g_d u_1 u_2 \cdots,
\quad\textrm{with}\quad
g_d \in GL(d,\reals)/SO(d),\quad u_k \in U_k.
\end{align}
The quotient by $SO(d)$ arises due to the symmetry of the internal metric $g_{mn}$: there are $\frac12d(d+1) = d^2-\frac12 d(d-1)$ scalar fields arising from the reduction of the metric. Another way of understanding the quotient is by considering the $D$-dimensional vielbein that has additional local Lorentz invariance. After dimensional reduction this is turned into a residual internal Lorentz symmetry $SO(d)$ that has to be fixed.

As explained in Section \ref{Rigidity}, and illustrated first for the examples of the $G_{2(2)}$ and $E_{8(8)}$ theories, the scalar manifold is a group manifold that is always isomorphic to the Borel subgroup $B$ of the candidate hidden symmetry group that can be read off the kinetic terms, independently of the precise values of the Chern-Simons couplings.\footnote{This is true in the neighbourhood of the critical values of the Chern-Simons couplings corresponding to the enhanced symmetry, and follows from theorems on deformation theory, see Section \ref{Rigidity}.  For large departures away from the critical values,  the scalar manifold might be given by a contraction of the Borel subgroup. Since our goal is to understand the enhancement of the symmetry when the Chern-Simons couplings take their critical values, we shall stay in the the neighbourhood of those critical values.}

The Borel subgroup $B \subset GL(d,\reals) \ltimes U$ acts simply transitively on the scalar manifold (i.e., on itself) by left multiplication. Invariant one-forms on the scalar manifold can be constructed from the Cartan--Maurer form
\begin{align}
\omega = V^{-1} d V
\end{align}
that takes values in the Lie algebra of the Borel  subgroup. The Lagrangian of the reduced theory can be expressed through these invariant one-forms and we will be interested in the symmetries of the metric on the scalar manifold expressed in this way. Any constant metric of the invariant one-forms realises the Borel symmetry. For generic choices of components of the constant metric, the isometry group will be just $B$ and its isotropy subgroup will be trivial.

The global symmetry $GL(d,\reals) \ltimes U$ acts also transitively on the scalar manifold but not simply transitively since the stability subgroup at any point is isomorphic to $SO(d)$. A constant metric of the invariant one-forms realises therefore the parabolic symmetry if it is invariant under $SO(d)$.   A symmetry enhancement of the hidden symmetry arises when the constant metric that is obtained after reduction of a specific choice of couplings in the higher-dimensional theory admits an even larger isotropy group. This will be illustrated in the examples in the next two sections. These two examples are (maximal) supergravity in $D=11$ and minimal supergravity in $D=5$. The two theories are both distinguished by possessing Chern--Simons couplings. We will keep this coupling as a parameter and study how it influences the properties of the global symmetry after reduction to three space-time dimensions.  We will then discuss more general cases.

We note that the choice of coordinates above is by no means unique and we have the freedom of performing field redefinitions on the scalar manifold. In particular, we can perform re-scalings of the fields and this will be important below.

\section{Enhancement for minimal $D=5$ supergravity and $G_{2(2)}$}
\label{G2sec}

We start by considering variations of minimal supergravity in $D=5$ and work in the conventions of~\cite{Compere:2009zh}. The bosonic Lagrangian density is given by
\begin{align}
\label{SUGRA5}
\mathcal{L}_{(5)} = R \star 1\!\! 1 - \frac12 \star F \wedge F + \frac{1}{3\sqrt{3}} F\wedge F \wedge A.
\end{align}
It is similar to~\eqref{SUGRA11} but now the gauge potential $A$ is a one-form and hence $F$ a two-form. It is known that the reduction of this theory to three dimensions exhibits a hidden $G_{2(2)}$ symmetry~\cite{Julia:1980gr,Mizoguchi:1998wv}.

\subsection{Parabolic global symmetry}

The dimensional reduction of~\eqref{SUGRA5} to three dimensions gives rise to a total of $8$ scalar fields that arise as follows:
\begin{itemize}
\item three scalar fields coming directly from the metric and are associated with $GL(2,\reals)/SO(2)$. Two out of the three scalars are of dilatonic type because they come from diagonal components of the metric. We will call the dilatons $\phi_1$ and $\phi_2$, the third metric scalar $\chi_1$
\item two scalars from the direct reduction of the one-form gauge potential that we call $\chi_2$ and $\chi_3$
\item one scalar from the reduction of the two-form gauge potential dual to $A$ in five dimensions that we call $\chi_4$
\item two scalars from dualising the two Kaluza--Klein vectors that we call $\chi_5$ and $\chi_6$
\end{itemize}
The notation for the scalar fields here is that of~\cite{Compere:2009zh}. An element of the scalar manifold can be written as
\begin{align}
V=e^{\frac{1}{2}\phi_1 h_1+\frac{1}{2}\phi_2 h_2}e^{\chi_1 e_1}e^{-\chi_2 e_2 + \chi_3 e_3}e^{\chi_6 e_6}e^{\chi_4 e_4-\chi_5 e_5},
\end{align}
where the $e_i$ are the shift generators of the parabolic global symmetry. The non-trivial commutators in their algebra are
\begin{align}
[e_1,e_2] &= e_3 ,&
[e_2,e_3] &= -\frac{2}{\sqrt 3} e_4,&
[e_2,e_4] &= -e_5,& 
[e_1,e_5] &= e_6 ,&
[e_3,e_4] &= -e_6.
\end{align}
The $h_i$ are the scaling generators of the dilatons and they commute with the $e_i$ according to 
\begin{align}
\lb k_1 h_1 + k_2 h_2, e_i\rb = (\vec{\alpha}_i \cdot \vec{k}) e_i ,
\end{align}
where
\begin{align}
\vec{\alpha}_1 = (-\sqrt{3},1), \quad\vec{\alpha}_2 = (\frac{2}{\sqrt 3},0)
\end{align}
and
\begin{align}
\vec{\alpha}_3=\vec{\alpha}_1+\vec{\alpha}_2,\; \vec{\alpha}_4=\vec{\alpha}_1 + 2\vec{\alpha}_2,\; \vec{\alpha}_5=\vec{\alpha}_1+3\vec{\alpha}_2,\; \vec{\alpha}_6=2\vec{\alpha}_1+3\vec{\alpha}_2 .
\end{align}
The generators $h_i$ and $e_i$ can be recognised as those of a Chevalley basis of $G_{2(2)}$. The vectors $\vec{\alpha}_i$ are the positive roots of $G_{2(2)}$, $\vec{\alpha}_1$ and $\vec{\alpha}_2$ being the simple ones.

The dimensionally reduced theory has the following metric on the scalar manifold:
\begin{align}
\label{scalars5}
ds^2=d\phi_1^2+d\phi_2^2+\sum_{i=1}^6 \omega_i^2 ,
\end{align}
where the invariant one-forms are given by
\begin{subequations}
\label{oneforms5}
\begin{align}
\omega_1 &= e^{\vec{\alpha}_1 \cdot \vec{\phi}/2}d\chi_1, \\
\omega_2 &= -e^{\vec{\alpha}_2 \cdot \vec{\phi}/2}d\chi_2, \\
\omega_3 &= e^{\vec{\alpha}_3 \cdot \vec{\phi}/2}\left( d\chi_3 - \chi_1 d\chi_2 \right), \\
\omega_4 &= e^{\vec{\alpha}_4 \cdot \vec{\phi}/2}\left( d\chi_4 + \frac{1}{\sqrt 3}(\chi_2 d\chi_3 - \chi_3 d\chi_2) \right), \\
\omega_5 &= -e^{\vec{\alpha}_5 \cdot \vec{\phi}/2}\left( d\chi_5 - \chi_2 d\chi_4 + \frac{1}{3 \sqrt 3}\chi_2(\chi_3 d\chi_2 - \chi_2 d\chi_3) \right), \\
\omega_6 &= e^{\vec{\alpha}_6 \cdot \vec{\phi}/2}\big( d\chi_6 - \chi_1 d\chi_5 + (\chi_1\chi_2 - \chi_3) d\chi_4, \\
&\qquad\qquad + \frac{1}{3 \sqrt 3}(-\chi_1\chi_2+\chi_3)(\chi_3 d\chi_2 - \chi_2 d\chi_3) \big),
\end{align}
\end{subequations}

The scalar manifold is isometric to the standard Borel subgroup of $G_{2(2)}$. We know that the symmetry spanned by $h_1$ and $e_1$ can be extended to also contain the lowering operator $f_1$ that completes the global symmetry to the (maximal) parabolic subgroup
\begin{align}
\label{G2para}
GL(2,\reals)\ltimes \left(\reals^2 + \reals^1 + \reals^2\right).
\end{align}

\subsection{Symmetry enhancement for minimal supergravity}

We now investigate additional hidden symmetries of the scalar metric~\eqref{scalars5} that has a very symmetric appearance. Since the scalar manifold is the Borel subgroup of $G_{2(2)}$ it is natural to investigate the action of the lowering operators of $f_i$ of $G_{2(2)}$. That $f_1$ is part of the parabolic symmetry was already argued above. In fact, it is more convenient to consider the compact generators $k_i=e_i-f_i$ instead of the lowering generators.  Since the scalar manifold is an homogeneous space (the Borel subalgebra acts transitively on it), we can also look at the variations at the special point $\phi=0$, $\chi=0$. 

It is sufficient to determine the action of $k_1$ and $k_2$ at zero; the other $k_i$ can be obtained from these by commutation. Performing the standard non-linear realisation one finds for $k_1$:
\begin{subequations}
\label{k13}
\begin{align}
\delta d\phi_1 &= \sqrt{3}\omega_1,&  \delta d\phi_2 &= -\omega_1 ,&\\
\delta \omega_1 &= -\sqrt{3} d\phi_1+d\phi_2,&  \delta \omega_2 &= \omega_3, & \delta \omega_3 &= -\omega_2,& \\
\delta \omega_4 &= 0,& \delta \omega_5 &= \omega_6,& \delta \omega_6 &= -\omega_5&
\end{align}
\end{subequations}
and for $k_2$:
\begin{subequations}
\label{k23}
\begin{align}
\delta d\phi_1 &= -\frac{2}{\sqrt{3}}\omega_2, &\delta d\phi_2 &= 0 ,&\\
\delta \omega_1 &= -\omega_3, & \delta \omega_2 &= \frac{2}{\sqrt{3}} d\phi_1,& \delta \omega_3 &= \omega_1 -\frac{2}{\sqrt{3}}\omega_4 ,&\\
\delta \omega_4 &= \frac{2}{\sqrt{3}}\omega_3-\omega_5, & \delta \omega_5 &= \omega_4,& \delta \omega_6& = 0.&
\end{align}
\end{subequations}
We note that $k_2$ mixes the groups $(\omega_1,\omega_2,\omega_3)$ and $(\omega_4,\omega_5,\omega_6)$, while $k_1$ does not. Plugging these explicit transformations into the coset metric~\eqref{scalars5} one checks that the transformations indeed leave the metric invariant. This recovers the well-known fact that minimal supergravity has a hidden global $G_{2(2)}$ symmetry.
 
\subsection{The role of the Chern--Simons coupling}
\label{CSG2}

Our main interest is to see the role of the Chern--Simons coupling. To this end we modify the Lagrangian~\eqref{SUGRA5} to
\begin{align}
\label{Lk5}
\mathcal{L}_\kappa = R \star 1\!\! 1 - \frac12 \star F \wedge F + \frac{\kappa}{3\sqrt{3}} F\wedge F \wedge A.
\end{align}
The value $\kappa=1$ corresponds to the theory considered above. The presence of $\kappa$ influences the scalar Lagrangian one obtains after reduction. The invariant forms are now 
\begin{subequations}
\label{invoneG2}
\begin{align}
\tilde{\omega}_1 &= e^{\vec{\alpha}_1 \cdot \vec{\phi}/2}d\chi_1, \\
\tilde{\omega}_2 &= -e^{\vec{\alpha}_2 \cdot \vec{\phi}/2}d\chi_2, \\
\tilde{\omega}_3 &= e^{\vec{\alpha}_3 \cdot \vec{\phi}/2}\left( d\chi_3 - \chi_1 d\chi_2 \right), \\
\tilde{\omega}_4 &= e^{\vec{\alpha}_4 \cdot \vec{\phi}/2}\left( d\chi_4 + \frac{\kappa}{\sqrt 3}(\chi_2 d\chi_3 - \chi_3 d\chi_2) \right), \\
\tilde{\omega}_5 &= -e^{\vec{\alpha}_5 \cdot \vec{\phi}/2}\left( d\chi_5 - \chi_2 d\chi_4 + \frac{\kappa}{3 \sqrt 3}\chi_2(\chi_3 d\chi_2 - \chi_2 d\chi_3) \right), \\
\tilde{\omega}_6 &= e^{\vec{\alpha}_6 \cdot \vec{\phi}/2}\big( d\chi_6 - \chi_1 d\chi_5 + (\chi_1\chi_2 - \chi_3) d\chi_4, \nn\\
&\qquad\qquad + \frac{\kappa}{3 \sqrt 3}(-\chi_1\chi_2+\chi_3)(\chi_3 d\chi_2 - \chi_2 d\chi_3) \big),
\end{align}
\end{subequations}
This has to be compared with~\eqref{oneforms5}. We see that only $\tilde{\omega}_4$, $\tilde{\omega}_5$ and $\tilde{\omega}_6$ differ. The non-linear shift invariances of these one-forms are
\begin{subequations}
\begin{align}
\chi_1 &\to \chi_1 + c_1,\\
\chi_2 &\to \chi_2 +c_2,\\
\chi_3 &\to \chi_3 +c_3 + c_1 \chi_2,\\
\chi_4 &\to \chi_4 +c_4 - \frac{\kappa}{\sqrt{3}}\left(c_2 \chi_3+(c_1c_2-c_3)\chi_2\right),\\
\chi_5 &\to \chi_5 +c_5+c_2\chi_4 - \frac{\kappa}{3\sqrt{3}}\left(2(c_1c_2^2-c_2c_3)\chi_2+2c_2^2\chi_3+(c_1c_2-c_3)\chi_2^2+c_2 \chi_2\chi_3\right),\\
\chi_6 &\to \chi_6 +c_6 + c_1\chi_5 + c_3 \chi_4\\
&\qquad-\frac{\kappa}{3\sqrt{3}} \left(2c_3(c_1 c_2-c_3)\chi_2  + 2 c_2c_3 \chi_3 +(2c_1c_2-c_3)\chi_2\chi_3  + c_1(c_1c_2-c_3)\chi_2^2 +c_2 \chi_3^2\right).\nn
\end{align}
\end{subequations}
This can be phrased more clearly in terms of commutators as follows. Let $E^a$ with $a=1,2$ be the generators of the first $\reals^2$ in~\eqref{G2para}, associated with the electric components $\chi_2$ and $\chi_3$ of the gauge potential. They transform as a doublet under the shift of the gravity scalar $\chi_1$
\begin{align}
\begin{pmatrix}1& c_1 \\0&1\end{pmatrix}
\begin{pmatrix} \chi_3 \\ \chi_2 \end{pmatrix} = \begin{pmatrix} \chi_3+ c_1 \chi_2\\ \chi_2 \end{pmatrix}
\end{align}
Under their own gauge transformations they simply transform by shifts $E^a$. These shifts commute according to
\begin{align}
\left[ E^a , E^b \right] = \frac{2\kappa}{\sqrt{3}} \epsilon^{ab} E,
\end{align}
where $E$ is the shift generator of the middle $\reals$ in~\eqref{G2para} and corresponds to the shift of the magnetic component of the gauge potential $\chi_4$. Letting $\tilde{E}^a$ be the shift symmetries of the remaining magnetic gravity components $\chi_5$ and $\chi_6$, one obtains the further commutator
\begin{align}
\left[ E, E^a \right] = \tilde{E}^a
\end{align}
that is independent of $\kappa$. All the shift generators transform with as standard tensor densities under $GL(2,\reals)$. 

We now return to the question of the symmetries of the reduced model. Starting from the $\kappa$-dependent Lagrangian~\eqref{Lk5} the reduced theory in three dimensions is expressed in terms of the invariant one-forms of~\eqref{invoneG2} as a non-linear sigma model with scalar metric given by
\begin{align}
ds^2=d\phi_1^2+d\phi_2^2+\sum_{i=1}^6 \tilde{\omega}_i^2 .
\end{align}
Rather than investigating the symmetries of this metric, we perform a field redefinition so that we can use the same results as before. As long as $\kappa\ne 0$, we can redefine
\begin{align}
\label{redG2}
\chi_4 \rightarrow \kappa \chi_4, \quad \chi_5 \rightarrow \kappa \chi_5, \quad \chi_6 \rightarrow \kappa \chi_6
\end{align}
to obtain
\begin{align}
\tilde{\omega}_i = \kappa \omega_i \quad\textrm{for $i=4,5,6$},
\end{align}
while the first three $\omega_i$ were identical already before. This means that the scalar metric of the $\kappa$-deformed model can be written as
\begin{align}
ds^2=d\phi_1^2+d\phi_2^2+\omega_1^2+\omega_2^2+\omega_3^2+\kappa^2(\omega_4^2+\omega_5^2+\omega_6^2)
\end{align}
Applying now the transformations $k_1$ and $k_2$ from equations~\eqref{k13} and~\eqref{k23} one finds that for $\kappa\neq 1$ only $k_1$ leaves this deformed metric invariant whereas $k_2$ does not. The $k_1$ symmetry has to be there because of the $GL(2,\reals)$ part of the parabolic symmetry that is always present. The enhancement due to the $k_2$ symmetry, however, is not present for values of the Chern--Simons coupling different from the value of minimal supergravity. This is the claimed result that the requirement of an enhanced symmetry implies the same constraints on the Chern--Simons coupling as supersymmetry would.

For the value $\kappa=0$ the Chern--Simons term is absent and the redefinition~\eqref{redG2} above is not allowed. The structure of the shift algebra simplifies to
\begin{align}
\left[ E^a, E^b \right]=0,\quad
\left[ E^a, E\right] =\tilde{E}^a.
\end{align}
(The remaining commutators are all zero.) This can be viewed as a contraction of the previous shift algebra.

A final comment concerns the enhanced symmetry that exists in pure $D=5$ gravity, i.e., without the gauge potential $A$. In this case it is known that there is an enhancement of the global symmetry to $SL(3,\reals)$. One might wonder whether at least this enhancement survives for arbitrary $\kappa$. Inspection of the relevant transformation shows that it is also broken because the matter fields fail to form an $SL(3,\reals)$ representation unless $\kappa=1$.

\section{Enhancement for $D=11$ supergravity and $E_{8(8)}$}
\label{E8sec}

We start with the following bosonic Lagrangian density in $D=11$
\begin{align}
\label{Lk11}
\mathcal{L}_\kappa = R \star 1\!\! 1 - \frac12 \star F \wedge F + \frac{\kappa}{6} F\wedge F \wedge A.
\end{align}
and follow the same procedure as in section \ref{G2sec}.
Compared to the bosonic part of $D=11$ supergravity in~\eqref{SUGRA11} we have introduced a free parameter $\kappa$ that controls the strength of the Chern--Simons self-interaction of the three-form $A$. The equation of motion for the three-form is
\begin{align}
\label{eom3}
d \star F = \kappa  F\wedge F = d( \kappa A \wedge F).
\end{align}
As shown, the equation of motion allows the reformulation in terms of the Bianchi identity of a six-form $\tilde{A}$ dual to the three-form $A$ such that
\begin{align}
\label{Adual}
d\tilde{A} = \star F- \kappa A \wedge F.
\end{align}
This equation is integrable by virtue of~\eqref{eom3}. The six-form $\tilde{A}$ has a five-form gauge parameter $\tilde\Lambda$ that leaves the original three-form $A$ unchanged. But the appearance of the naked $A$ on the right-hand side of~\eqref{Adual} implies that this equation is only gauge-invariant if $\tilde{A}$ transforms non-trivially under the gauge parameter $\Lambda$ of the three-form. The full set of matter gauge transformations are
\begin{align}
\label{gauge11}
A &\to A + d\Lambda,\nn\\
\tilde{A} &\to \tilde{A} + d\tilde{\Lambda} -\kappa \Lambda\wedge F. 
\end{align}
Morally, the last term should be thought of as integrated by parts to look more like $d\Lambda \wedge A$ and it indicates a non-trivial commutator of the shift symmetries associated with the gauge symmetries of the three-form.

In addition, there are 8 scalar components in three dimensions from the dualisation of the eight Kaluza--Klein vectors in the metric sector.

\subsection{Parabolic global symmetries}

Dimensional reduction of~\eqref{Lk11} to three space-time dimensions therefore gives rise to the following global symmetries
\begin{align}
GL(8,\reals) \ltimes\left( \reals^{56} + \reals^{28} + \reals^{8}\right),
\end{align}
where the $56$ shifts come from the direct scalars from the three-form, the $28$ shifts from the scalars from the dual six-form and the $8$ shifts from the dual graviton.  The structure of the matter shift symmetries can be read off from the gauge transformations ~\eqref{gauge11} in $D=11$ that give rise to the shifts in three space-time dimensions. The presence of the Chern--Simons coupling implies that
\begin{align}
\lb E^{a_1a_2a_3} , E^{a_4a_5a_6} \rb = \kappa E^{a_1a_2a_3a_4a_5a_6}
\end{align}
as anticipated in~\eqref{336} but where now the indices can take eight different values. As long as $\kappa\neq 0$, we can introduce a re-scaled shift generator $\tilde{E}^{a_1\ldots a_5a_6}= \kappa E^{a_1\ldots a_5a_6}$ that brings the above commutation relation into a standard form. We also see that for $\kappa=0$, when there is no self-interaction of the three-form, the gauge shift symmetries abelianise.

Analogously to the algebra of the shift symmetries discussed in section~\ref{CSG2}, the shift symmetries of the six-form and of the three-form close into shifts of the dualised graviphotons according to
\begin{align}
\lb E^{m_1m_2m_3}, E^{m_4\ldots m_9} \rb = E^{m_1m_2m_3 [m_4\ldots m_8|m_9]} = \epsilon^{m_1m_2m_3 [m_4\ldots m_8} E^{m_9]} 
\end{align}
where we have written the $8$ shift symmetries also in terms of a mixed hook Young tableaux as is more customary for large $E$-type symmetries~\cite{West:2001as,Damour:2002cu}. This commutator does not depend on the value of the Chern--Simons coupling $\kappa$. 

The scalar manifold is the Borel subgroup $B(E_{8(8)})$ of $E_{8(8)}$ and the reduced scalar metric can be written in terms of  one-forms invariant under the shift symmetries. The result is 
\begin{align}
ds^2 = \sum_{i=1}^8 d\phi^2_i + \sum_{a<b} \tilde{\omega}_{ab}^2 + \sum_{a<b<c} \tilde{\omega}_{abc}^2 + \sum_{a_1<\ldots<a_6} \tilde{\omega}_{a_1\ldots a_6}^2 + \sum_{a} \tilde{\omega}_a^2.
\end{align}
The first two terms correspond to the Borel subgroup of $GL(8,\reals)$ and the last three-terms to the unipotent shift symmetries. The tilde indicates that the invariant one-forms have been computed in the normalisation that follows from the reduction of the theory with arbitrary Chern--Simons coupling $\kappa$. Explicit formulae can be found in \cite{Cremmer:1997ct}.

\subsection{Symmetry enhancement for the correct Chern--Simons coupling}

For $\kappa\neq 0$ one can bring the Borel shifts into canonical $E_{8(8)}$ form at the expense of changing the normalisation of some of the generators. In terms of the invariant one-forms $\tilde{\omega}$ this means that we change to the canonically normalised one-forms $\omega$. The metric on the scalar manifold then becomes
\begin{align}
ds^2 = \sum_{i=1}^8 d\phi^2_i + \sum_{a<b} \omega_{ab}^2 + \sum_{a<b<c} \omega_{abc}^2 + \kappa^2\left(\sum_{a_1<\ldots<a_6} \omega_{a_1\ldots a_6}^2 + \sum_{a} \omega_a^2\right).
\end{align}
For the value $\kappa^2=1$, this metric has an additional $SO(16)$ symmetry that acts on the $128$ scalars in the spinor representation. This corresponds to the symmetry enhancement from the parabolic shift symmetry to the full $E_{8(8)}$ hidden symmetry of Cremmer and Julia.

For values of $\kappa\neq \pm 1$, the symmetry is reduced and one only has the $SO(8)$ isotropy acting on the tensors in the corresponding tensorial representation. For $\kappa=0$ the Borel shift symmetry is contracted.

\section{Scalar manifold and rigidity theorems}
\label{Rigidity}
We have seen in the previous examples that except for $\kappa=0$, one could always absorb $\kappa$ through redefinitions in the structure constants of the smaller symmetry algebra.  This smaller symmetry algebra present for all values of $\kappa$ has therefore always the same structure provided that $\kappa$ does not vanish. This is not an accident as we now explain.

The situation is the following. Consider a theory in $D$ spacetime dimensions that has hidden symmetry algebra $E$ (which can be any simple Lie algebra \cite{Breitenlohner:1987dg,Cremmer:1999du}) upon dimensional reduction to three dimensions.  This theory involves $p$-form fields and Chern-Simons couplings.  For the critical values of the Chern-Simons coefficient for which $E$ appears, the complete scalar manifold in three dimensions can be identified with the group manifold of the Borel subgroup $B(E)$, which is part of the symmetry\footnote{We use here the same letters for the Lie algebra and the corresponding group.  No confusion should arise since the context is clear.}.   When deforming away from the critical point, the symmetry algebra is reduced and contains, as we have seen,  $B(GL(d,\reals)) \ltimes U$, which has same dimension as $B(E)$. The structure constants of the subalgebra $B(GL(d,\reals)) \ltimes U$ depend continuously on the Chern-Simons coefficients, and so $B(GL(d,\reals)) \ltimes U$ is a deformation of $B(E)$.  But by the rigidity theorems of \cite{LegerLuks}, the algebra $B(E)$ admits no non-trivial deformation.  Hence, $B(GL(d,\reals)) \ltimes U$ is isomorphic with $B(E)$.

Of course the argument is valid in the vicinity of the critical values and does not eliminate the possibility of having contractions of $B(E)$ under deformations going out of that vicinity, just as the rigidity of simple Lie algebras does not eliminate the possibility to contract them to abelian algebras.

\subsection*{Acknowledgements}
M.H. and V.L. would like to thank Axel Kleinschmidt for his kind hospitality in the Max Planck Institute, where dicussions leading to this work began. M.H. thanks the Alexander von Humboldt Foundation for a Humboldt Research Award.  The work of M.H. and V. L. is partially funded by the ERC through the ``SyDuGraM\textquotedblright{}\ Advanced Grant,  by FNRS-Belgium (convention
FRFC PDR T.1025.14 and convention IISN  4.4503.15) and  by the ``Communaut\'e Fran\c{c}aise
de Belgique\textquotedblright{}\ through the ARC program.


\begin{thebibliography}{20}

\bibitem{Cremmer:1979up}
  E.~Cremmer and B.~Julia,
  ``The SO(8) Supergravity,''
  Nucl.\ Phys.\ B {\bf 159} (1979) 141.
  %%CITATION = NUPHA,B159,141;%%
  
 \bibitem{LegerLuks} G. Leger and E. Luks,
 ``Cohomology theorems for Borel-like solvable Lie algebras in arbitrary characteristic",
 Can. J. Math. {\bf  XXIV} (1972) 1019.
  
\bibitem{Cremmer:1978km}
  E.~Cremmer, B.~Julia and J.~Scherk,
  ``Supergravity Theory in Eleven-Dimensions,''
  Phys.\ Lett.\ B {\bf 76} (1978) 409.
  %%CITATION = PHLTA,B76,409;%%

\bibitem{Bern:2011qn}
  Z.~Bern, J.~J.~Carrasco, L.~J.~Dixon, H.~Johansson and R.~Roiban,
  ``Amplitudes and Ultraviolet Behavior of N = 8 Supergravity,''
  Fortsch.\ Phys.\  {\bf 59} (2011) 561
  [arXiv:1103.1848 [hep-th]].
  %%CITATION = ARXIV:1103.1848;%%

\bibitem{Beisert:2010jx}
  N.~Beisert, H.~Elvang, D.~Z.~Freedman, M.~Kiermaier, A.~Morales and S.~Stieberger,
  ``$E_{7(7)}$ constraints on counterterms in N=8 supergravity,''
  Phys.\ Lett.\ B {\bf 694} (2010) 265
  [arXiv:1009.1643 [hep-th]].
  %%CITATION = ARXIV:1009.1643;%%

\bibitem{Kallosh:2011qt}
  R.~Kallosh,
  ``N=8 Counterterms and $E_{7(7)}$ Current Conservation,''
  JHEP {\bf 1106} (2011) 073
  [arXiv:1104.5480 [hep-th]].
  %%CITATION = ARXIV:1104.5480;%%

\bibitem{Bossard:2011ij}
  G.~Bossard and H.~Nicolai,
  ``Counterterms vs. Dualities,''
  JHEP {\bf 1108} (2011) 074
  [arXiv:1105.1273 [hep-th]].
  %%CITATION = ARXIV:1105.1273;%%

\bibitem{Gunaydin:2013pma}
  M.~Gunaydin and R.~Kallosh,
  ``Obstruction to $E_{7(7)}$ Deformation in N=8 Supergravity,''
  arXiv:1303.3540 [hep-th].
  %%CITATION = ARXIV:1303.3540;%%
 
\bibitem{Julia:1980gr}
  B.~Julia,
  ``Group Disintegrations,''
  Conf.\ Proc.\ C {\bf 8006162} (1980) 331.
  %%CITATION = CONFP,C8006162,331;%% 

\bibitem{Lu:1995yn}
  H.~Lu and C.~N.~Pope,
  ``P-brane solitons in maximal supergravities,''
  Nucl.\ Phys.\ B {\bf 465} (1996) 127
  [hep-th/9512012].
  %%CITATION = HEP-TH/9512012;%%

\bibitem{Cremmer:1997ct}
  E.~Cremmer, B.~Julia, H.~Lu and C.~N.~Pope,
  ``Dualization of dualities. 1.,''
  Nucl.\ Phys.\ B {\bf 523} (1998) 73
  [hep-th/9710119].
  %%CITATION = HEP-TH/9710119;%%

\bibitem{Lambert:2001he}
  N.~D.~Lambert and P.~C.~West,
  ``Coset symmetries in dimensionally reduced bosonic string theory,''
  Nucl.\ Phys.\ B {\bf 615} (2001) 117
  [hep-th/0107209].
  %%CITATION = HEP-TH/0107209;%%

\bibitem{Pope}
 C.~N.~Pope,
 ``Kaluza--Klein theory,'' \href{http://people.physics.tamu.edu/pope/ihplec.pdf}{
 http://people.physics.tamu.edu/pope/ihplec.pdf} [accessed on 3/5/2015]

\bibitem{Ehlers}
  J. Ehlers,
   ``Transformations of Static Exterior Solutions of Einstein’s Gravitational Field Equations into Different Solutions by Means of Conformal Mappings,'' {\sl Les th\'eories physiques de la gravitation}, CNRS, Paris, 1959.

\bibitem{West:2001as}
  P.~C.~West,
  ``E(11) and M theory,''
  Class.\ Quant.\ Grav.\  {\bf 18} (2001) 4443
  [hep-th/0104081].
  %%CITATION = HEP-TH/0104081;%%
  
\bibitem{Damour:2002cu}
  T.~Damour, M.~Henneaux and H.~Nicolai,
  ``E(10) and a 'small tension expansion' of M theory,''
  Phys.\ Rev.\ Lett.\  {\bf 89} (2002) 221601
  [hep-th/0207267].
  %%CITATION = HEP-TH/0207267;%%  

\bibitem{Compere:2009zh}
  G.~Compere, S.~de Buyl, E.~Jamsin and A.~Virmani,
  ``G2 Dualities in D=5 Supergravity and Black Strings,''
  Class.\ Quant.\ Grav.\  {\bf 26} (2009) 125016
  [arXiv:0903.1645 [hep-th]].
  %%CITATION = ARXIV:0903.1645;%%

\bibitem{Mizoguchi:1998wv}
  S.~Mizoguchi and N.~Ohta,
  ``More on the similarity between D = 5 simple supergravity and M theory,''
  Phys.\ Lett.\ B {\bf 441} (1998) 123
  [hep-th/9807111].
  %%CITATION = HEP-TH/9807111;%%
 
\bibitem{Breitenlohner:1987dg}
  P.~Breitenlohner, D.~Maison and G.~W.~Gibbons,
  ``Four-Dimensional Black Holes from Kaluza-Klein Theories,''
  Commun.\ Math.\ Phys.\  {\bf 120} (1988) 295.
  %%CITATION = CMPHA,120,295;%% 
 
 %\cite{Cremmer:1999du}
\bibitem{Cremmer:1999du}
  E.~Cremmer, B.~Julia, H.~Lu and C.~N.~Pope,
  ``Higher dimensional origin of D = 3 coset symmetries,''
  hep-th/9909099.
  %%CITATION = HEP-TH/9909099;%%
   
\end{thebibliography}
\end{document}